\documentstyle[twocolumn,prl,aps,epsf,floats]{revtex}

\begin{document}
\draft
\twocolumn[\hsize\textwidth\columnwidth\hsize\csname@twocolumnfalse\endcsname
\title{Antiferromagnetic interlayer exchange coupling\\ across an amorphous
metallic spacer layer}
\draft
\author{D.E.~B\"urgler\cite{deb}$^1$, D.M.~Schaller$^1$, C.M.~Schmidt$^1$,
F.~Meisinger$^1$,\\ J.~Kroha$^2$, J.~McCord$^3$, 
A.~Hubert$^3$, and H.-J.~G\"untherodt$^1$}
\address{$^1$Institut f\"ur Physik, Universit\"at
Basel, Klingelbergstrasse 82, CH-4056 Basel, Switzerland}
\address{$^2$Institut f\"ur Theorie der Kondensierten Materie, 
Universit\"at Karlsruhe, Postfach 6980, D-76128 Karlsruhe, Germany}
\address{$^3$Institut f\"ur Werkstoffwissenschaften 6,
Universit\"at Erlangen-N\"urnberg, Martensstr. 7, D-91058 Erlangen,
Germany}
\date{\today}
\maketitle
\begin{abstract}
By means of magneto-optical Kerr effect we observe for the first time 
antiferromagnetic coupling between ferromagnetic layers across an 
amorphous metallic spacer layer.  Biquadratic coupling occurs at 
the transition from a ferromagnetically to an antiferromagnetically 
coupled region.  Scanning tunneling microscopy images of all involved 
layers are used to extract thickness fluctuations and to verify the 
amorphous state of the spacer.  The observed antiferromagnetic 
coupling behavior is explained by RKKY interaction taking into account 
the amorphous structure of the spacer material.
\end{abstract}
]
    
\narrowtext
The magnetic exchange coupling between two ferromagnetic layers across 
a metallic spacer has recently attracted considerable experimental and 
theoretical interest\cite{bland2}.  Oscillatory exchange coupling with 
the alignment of the magnetization vectors alternating between 
parallel (ferromagnetic, FM) and antiparallel (antiferromagnetic, AFM) 
with increasing spacer thickness was found for most 
transition-metal\cite{park91-1} and noble-metal \cite{leng94-1} and 
also for some alloy\cite{brun97-1} spacers.  Theoretically, the 
oscillating behavior has been explained by the interplay between the 
Ruderman-Kittel-Kasuya-Yosida (RKKY) interaction and the discrete 
spacer thickness\cite{coeh91-1,chap91-1}.  Assuming a spherical Fermi 
surface with Fermi wave number $k_{F}$ and reducing the RKKY wave 
number $2k_{F}$ to the first Brillouin zone of the planar periodic 
structure with a lattice constant $d$, the oscillation period 
$\Lambda$ is given by $ 1/\Lambda = |1/\lambda - n/d|,\ n=1,2,\dots$, 
with $\lambda = \pi/k_{F}$.  The same relation for $\Lambda$ can 
be derived in a picture where the oscillatory exchange interaction 
with wave length $\lambda = \pi/k_{F}$ originates from spin-dependent 
quantum well states in the spacer\cite{edwa91-1,orte92-1} instead of 
RKKY interaction.  Phenomenologically, the FM and AFM coupling is 
described by a bilinear energy density term 
$-J_{1}(z)\cos(\vartheta)$, where $\vartheta$ is the angle between the 
magnetizations of the two ferromagnetic layers and $z$ the spacer 
thickness.  In addition, a contribution favoring perpendicular 
arrangement of the magnetizations (90$^{\circ}$ coupling) has been 
observed\cite{RUEH91-1}.  It is parameterized by a biquadratic energy 
density term, $-J_{2}(z)\cos^2(\vartheta)$.  Several  
models for the biquadratic coupling have been proposed
\cite{slon95-1}.  
Refs.~\cite{SLON91-1,DEMO94-1} 
relate this type of coupling to 
thickness fluctuations of the spacer originating from interface 
roughness.

Amorphous spacers provide the possibility to study the interlayer 
coupling in the absence of the structural discreteness which plays a 
crucial role in all theoretical models proposed so far.  A previous 
study employing amorphous {\em semiconducting} spacers\cite{brin94-1} 
revealed AFM coupling with a positive temperature coefficient, which 
was interpreted as resonant tunneling of polarized electrons through 
defect-generated localized states in the gap of the semiconducting 
spacer.  Hence, a comparison to the mentioned theoretical models 
derived for conduction electrons was not possible.  Fuchs {\em et 
al.}\cite{fuch97-1} very recently investigated amorphous metallic AuSn 
spacers and found FM and 90$^{\circ}$ coupling originating from 
dipolar interactions while the exchange coupling is strongly 
suppressed.  AFM coupling was not observed.  In this Letter we show 
for the first time that also an {\em amorphous metallic} spacer can 
mediate AFM coupling between ferromagnetic layers and present a model 
to explain the findings.

\begin{figure*}[t]    
	\epsffile{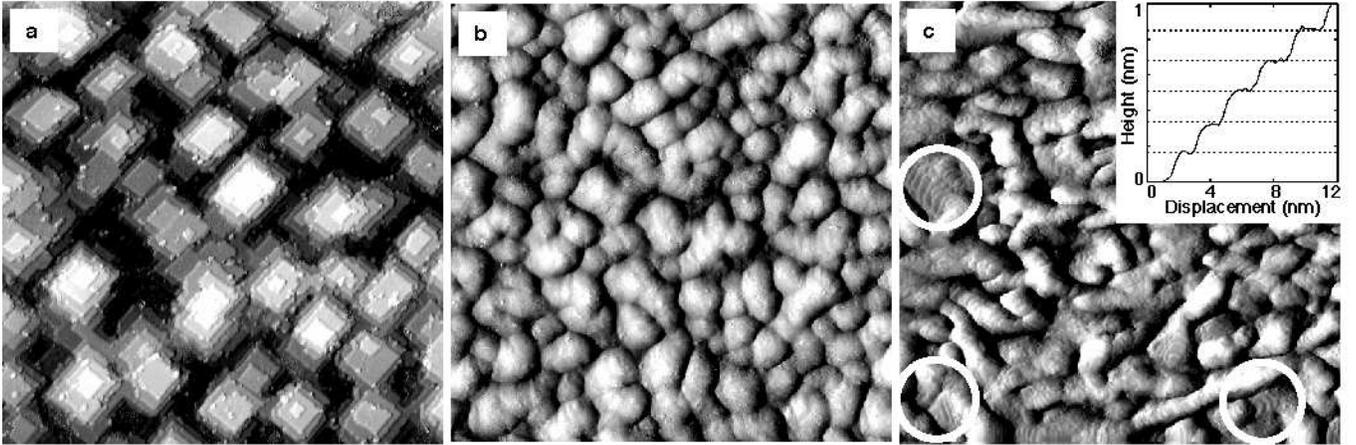} 
	\caption{ STM images (100$\times$100~nm$^2$) 
	of the layers forming the sandwich structure: 
	(a) 5~nm singlecrystalline bottom Fe layer (vertical range: 1~nm).  
	(b) 2~nm $a$-CuZr layer grown on the Fe layer shown in (a) (vertical 
	range: 2~nm).  
	(c) 5~nm polycrystalline top Fe layer grown on the 
	$a$-CuZr layer shown in (b) (vertical range: 3~nm).  Rings 
	indicate some stepped areas and the inset shows a series of 
	equally high steps.}
	\label{f-stm}
\end{figure*}
Sample preparation and all measurements with the exception of Kerr 
microscopy are performed in an ultra-high vacuum (UHV) system with a 
base pressure of 5$\times$10$^{-11}$~mbar which is equipped with an 
$e$-beam deposition system, scanning tunneling microscopy (STM), 
low-energy electron diffraction, Auger and X-ray photoemission 
electron spectroscopy (AES, XPS), and a magneto-optical Kerr effect 
(MOKE) setup.  Amorphous metallic Cu$_{x}$Zr$_{100-x}$ ($a$-CuZr) is 
used as spacer material.  This alloy vapor-quenches in the amorphous 
state and stays amorphous at temperatures up to 600~K for the 
composition used in this work, $x\approx 65 \%$ \cite{rege90-1}.   
We grow the $a$-CuZr spacers by coevaporation from two $e$-beam 
sources onto substrates held at 490~K.  The deposition rates of Cu and 
Zr are individually controlled by two quartz thickness monitors.   
Wedge-shaped spacers with a slope of 0.5~nm/mm are grown by linearly 
moving a shutter in front of the substrate during deposition.  The 
$a$-CuZr spacer is sandwiched by 5~nm thick Fe layers.  The bottom one 
is epitaxially grown in (001) orientation on an Ag(001)/Fe/GaAs(001) 
substrate following the optimized growth procedures described in 
Refs.~\cite{buer97-1,buer96-1}.  The top Fe layer is grown at 300~K 
and adopts a polycrystalline structure.  The composition and the 
cleanness of all layers are confirmed by XPS and AES.  For the {\em 
ex-situ} Kerr microscopy analysis the samples are coated with 5~nm Ag 
and with a ZnS layer for enhancement of the magneto-optical contrast.  
MOKE-measurements in UHV before and after coating with Ag do not show 
any effect of the cap layer.

Figure~\ref{f-stm}a depicts an STM image of the surface of the 
epitaxial bottom Fe layer.  The morphology is characterized by a 
regular arrangement of equally sized quadratic table mountains.  They 
are delimited by series of single-atomic steps running along 
$\langle100\rangle$ axes.  The rms-roughness 
$\sigma=\sqrt{\langle{z^2}\rangle}$ amounts to $\sigma_{Fe} = $ 
0.21~nm.  The STM image (Fig.~\ref{f-stm}b) of the surface of a 2~nm 
thick $a$-CuZr film, in contrast, shows the typical appearance of an 
amorphous vapor-quenched thin film: an irregular arrangement of growth 
hillocks.  $\sigma_{CuZr}$ amounts to 0.44~nm.  The weak finestructure 
consists of nm-sized irregularly arranged features very similar to 
previously published STM images of sputtered or laser-quenched 
amorphous ribbons \cite{SCHA96-1}.  Our findings are in good 
agreement with recent STM results of vapor-quenched $a$-ZrAlCu 
thin films \cite{rein97-1}.  The STM image of the top Fe layer 
(Fig.~\ref{f-stm}c) reveals a grain structure with many single-atomic 
steps (inset) proving its polycrystalline state.

\begin{figure}[tb]
    \hspace*{-6mm}    
    \epsffile{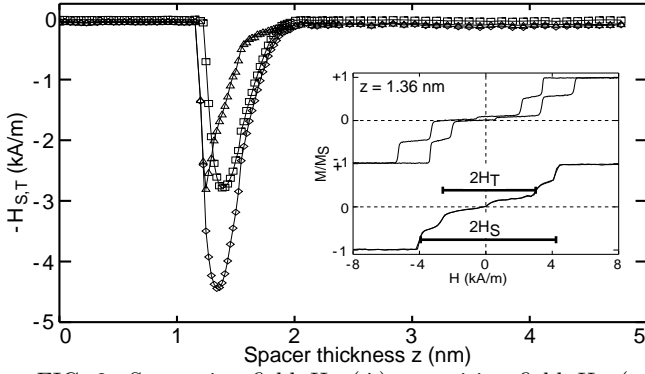}
    \caption{Saturation field $H_{S}$ ($\Diamond$), transition field 
    $H_{T}$ ($\Box$), $H_{S} - H_{T} = 2J_{2}/c$ ($\triangle$) as a 
    function of spacer thickness $z$.  Inset: Hysteretic and anhysteretic 
    magnetization curve taken at $z=1.36$~nm.}
\label{f-moke}
\end{figure}

Longitudinal MOKE is used to record magnetization curves.  The 
external field is applied parallel to a $[100]$ easy axis of the 
bottom Fe layer.  A magnetization curve in units of the saturation 
magnetization $M_{S}$ taken at $z = 1.36$~nm is shown in the upper 
part of the inset in Fig.~\ref{f-moke}.  Three plateaus at 0 and 
$\pm$0.5~$M_{S}$ indicate one phase with vanishing net magnetization 
and two phases with contribution from only one Fe layer, respectively.  
AFM coupling at zero field and perpendicular orientation of the 
magnetizations with one of them parallel to the external field at 
intermediate field strength is compatible with this magnetization 
curve.  Therefore we call these plateaus {\it AFM} and {\it 
90$^{\circ}$ plateau}, respectively.  The small step close to $H=0$ 
originates from the finite sampling depth of the light and from a weak 
inequality in thickness or saturation magnetization of the two Fe 
layers.  In order to exclude coercive effects, which could cause a 
plateau at $M = 0$ even in the case of decoupled layers due to 
different coercive fields, anhysteretic magnetization curves 
\cite{SCHA94-3} are measured.  This is achieved by superimposing to 
the static external field a decaying AC magnetic field prior to the 
measurement of each data point of the magnetization curve.  The 
coincidence of forward and backward scan in the lower curve of the 
inset in Fig.~\ref{f-moke} confirms the anhysteretic measurement mode.  
Obviously, the three plateaus are still visible although their edges 
are now rounded indicating a more continuous rotation of the 
magnetizations.  The width of the AFM plateau is almost unchanged, 
whereas the inequality of the width of the 90$^{\circ}$ plateaus has 
disappeared.

In order to quantify the magnetization curves we define the saturation 
field $H_{S}$ and the transition field $H_{T}$ between the AFM and 
90$^{\circ}$ plateaus: $H_{S}$ ($H_{T}$) is half the field interval 
between the values where $M(H) = \pm 0.75 M_{S}$ ($\pm 0.25 M_{S}$) 
averaged over forward and backward scan.  Figure~\ref{f-moke} shows 
the dependence of $H_{S}$ ($\Diamond$) and $H_{T}$ ($\Box$) on the 
spacer thickness $z$.  Both show a pronounced peak at $z = 1.36$~nm 
and vanish for $z > 2.05$~nm.  Note that the onset of the peak of 
$H_{T}(z)$ is shifted by 0.07~nm towards larger $z$ values with 
respect to onset of $H_{S}(z)$ at $z = 1.15$~nm.  We estimate the 
total coupling strength as $J_{1}(z) + J_{ 2}(z) = c 
H_{S}(z)$.  $c = -\frac{\mu_{0}}{2}(m_{1}d_{1} + m_{2}d_{2})$, where 
$m_{i}$ and $d_{i}$ denote the saturation magnetization and the 
thickness of the two Fe layers, respectively.  Using the Fe bulk value 
$m_{1,2} = 1.714\times 10^{6}$~A/m and $d_{1,2} = 5$~nm we obtain a 
maximum coupling strength of -0.05~mJm$^{-2}$.  This value is in the 
range found for transition-metal and noble-metal 
spacers\cite{park91-1,leng94-1}.  $J_{2}$ is proportional to the 
width of the 90$^{\circ}$ plateaus and therefore $J_{2}(z)=\frac{c}{2} 
(H_{S}(z) - H_{T}(z))$ ($\triangle$ in Fig.~\ref{f-moke}).  It is 
clearly seen that the biquadratic coupling is strongest at the onset 
of the AFM coupling.

\begin{figure}[tb]
	\epsffile{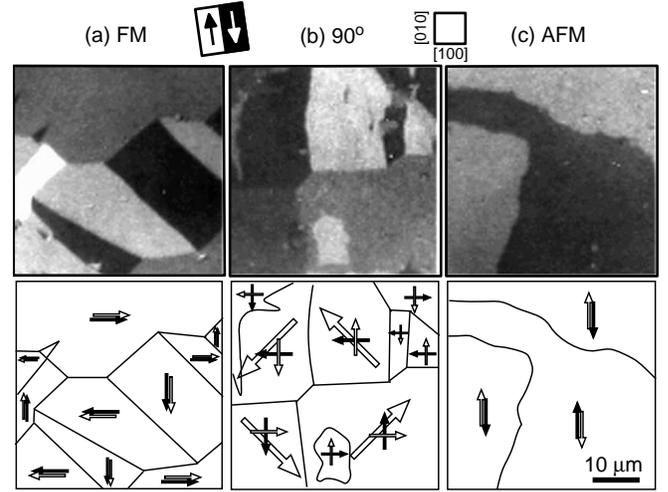} 
	\caption{Kerr microscopy domain pattern images 
	(43$ \times$ 43~$\mu$m$^2$) showing FM coupling at $z=1.06$ nm 
	(a), 90$^{\circ}$ coupling at $z=1.15$ nm (b), and AFM coupling at 
	$z=1.25$ nm (c).  The magneto-optical sensitivity axis is slightly 
	rotated with respect to the easy axes in order to obtain contrast 
	differences for domains with horizontal but opposite magnetization 
	components.  Open (filled) arrows represent the direction of the 
	magnetization of the top (bottom) Fe layer.  The resulting net 
	magnetization in (b) is shown by larger arrows.}
\label{f-km}
\end{figure}

The validity of this interpretation is confirmed by direct domain 
observation using Kerr microscopy\cite{schm85-1}.  
Figures~\ref{f-km}a-c show the domain patterns in the demagnetized, 
field-free state at three different positions along the wedge 
corresponding $z=$ 1.06~nm, 1.15~nm, and 1.25~nm.  A domain 
configuration of FM coupled films with predominantly 90$^{\circ}$ and 
180$^{\circ}$ domain walls and the magnetizations oriented parallel to 
the easy $\langle100\rangle$ axes of the bottom Fe layer is visible in 
Fig.~\ref{f-km}a.  In the transition region (Fig.~\ref{f-km}b) all 
characteristics of 90$^{\circ}$ coupled layers\cite{RUEH91-1,SCHA95-3} 
are observed.  The straight domain walls are rotated by 45$^{\circ}$ 
as compared to the FM coupled region separating areas with differently 
oriented net magnetization (large arrows).  Irregularly shaped walls 
occur between domains with the same net magnetization.  Domain 
observations reveal a width of the transition region of 140~$\mu$m 
corresponding to $\Delta z=$ 0.06~nm in good agreement with the 
shifted onset of $H_{T} (\Box)$ compared to $H_{S} (\Diamond)$ in 
Fig.~\ref{f-moke}.  Figure~\ref{f-km}c reveals exclusively irregularly 
shaped domain walls originating from AFM 
coupling\cite{RUEH91-1,SCHA95-3}.  For $z\buildrel{>}\over{\sim}$ 2~nm 
the domain pattern indicate weak FM coupling.

The appearance of a single AFM minimum of the coupling at $z=$ 1.36~nm 
(Fig.~\ref{f-moke}) may be explained in terms of RKKY interaction in 
combination with the oscillatory ion density correlations with wave 
number $k_p$ in the amorphous spacer.  The latter imply that spacer 
thicknesses of $z=2\pi n /k_p$, $n=0,1,2,\dots $, are preferred in the 
deposition process and thus, lead to oscillations of the interlayer 
coupling with wave number $|2k_F - k_p|$, in analogy to crystalline 
spacers \cite{coeh91-1,chap91-1}.  However, the rapid decay of the 
structural correlations in amorphous materials strongly damps these 
oscillations, so that only the first AFM minimum is observable.  In 
order to quantify this model, we calculate the bilinear coupling 
energy density $J_1(z)$ between the Fe layers, which is an average over 
the fluctuating spacer thickness,
\begin{equation}
J_1(z)=\int _0^{\infty} K(z'){\cal P}_z(z') dz'.
\label{e-j}
\end{equation}  
Here, $K (z')$ is the RKKY coupling between the Fe layers in a planar 
geometry for a fixed spacing $z'$ and ${\cal P}_{z}(z')$ is the 
probability for the occurrence of a spacer thickness $z'$, when the 
average thickness is $z$.  ${\cal P}_{z}(z')$ may be written in terms 
of the layer correlation function $G(z')$ as ${\cal P}{_z}(z') = 
G(z')\exp(-(z'-z)^2/\Delta)/\sqrt{\pi\Delta}$.  $\Delta$ is a measure 
for the spacer thickness fluctuations.  Assuming uncorrelated 
roughness an upper limit for $\Delta$ can be calculated from the rms 
roughnesses $\sigma$ of the interfaces, $\Delta = 
2(\sigma_{CuZr}^2+\sigma_{Fe}^2) \simeq (0.69$~nm$)^2$.  In the planar 
geometry $G(z')$ is the 1-dimensional Fourier transform of the 
amorphous structure factor $S(|\vec q |)$.
\begin{figure}[tb]
	\hspace*{-6mm} 
	\epsffile{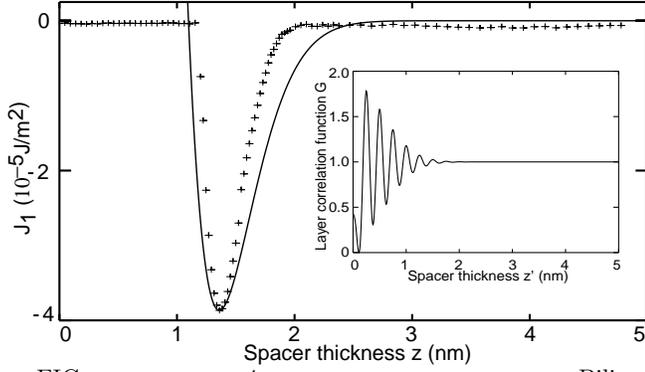} \caption{Bilinear part of the 
	interlayer coupling for an $a$-Cu$_{65}$Zr$_{35}$ spacer as 
	calculated from Eqs.  (1--3) (solid line) and as determined 
	experimentally by $J_{1}(z) = \frac{c}{2} (H_{S}(z) + 
	H_{T}(z))$ (+).  Inset: layer correlation function $G(z')$ of the 
	amorphous material.  $k_{F}$ and $k_{p}$ are determined from a 
	free electron model and from the average ion density of 
	$a$-Cu$_{65}$Zr$_{35}$, respectively, with Cu and Zr valences 
	$Z_{Cu} = 1$ and $Z_{Zr} = 2$.  The parameter values used are $w = 
	0.1k_p$, $A = 3$, $z_{o}=1.45 a/2 =0.208$~nm, and $V_1=V_2\equiv 
	V$.  The constant $|V|^2N(0)$ is fitted to account for the 
	measured size of the effect and agrees qualitatively with the 
	value estimated from the Fe Curie temperature.}
\label{f-model}
\end{figure}
It has the typical form \cite{haeu92-1}
\begin{equation} 
   S(|\vec q |) = 1 - {\rm e}^{-(|\vec q |/k_p)^2} + 
   A {\rm e}^{-\bigl( \frac{|\vec q |-k_p}{w} \bigr)^2},
\label{e-s}
\end{equation}  
showing a pronounced peak at $q=k_p$ and approaching a constant value 
for $q\gg k_p$.  
We thus obtain the layer correlation 
function shown in the inset of Fig.~\ref{f-model} exhibiting typical 
damped oscillatory behavior.

Considering purely bilinear coupling the localized moments in each Fe 
film are rigidly aligned parallel to each other.  Through an exchange 
coupling $V_1$, each Fe atom induces an oscillating polarization of 
the surrounding electron sea $\propto \cos(2k_Fr - \varphi)/r^3$, 
which is transferred to the atoms of the other Fe layer through the 
paramagnetic spacer.  We allow for a phase shift $\varphi$ of the RKKY 
oscillations.  In addition, the Fe local moments at the layer surfaces 
couple directly to the electron states of the $a$-CuZr layer via an 
exchange coupling $V_2$.  The resulting $K(z)$ for a fixed spacing $z$ 
then reads ($z>\pi/k_F$)
\begin{eqnarray}  
K(z)=&-&2\pi N(0)\biggl[\biggl(\frac{|V_2|^2}{(2k_Fa)^2}+
\frac{4|V_1|^2}{(2k_Fa)^4}\biggr)\frac{{\rm sin}(\zeta-\varphi)}
{\zeta^2}\nonumber\\
&+&\frac{4{\rm Re}(V_1V_2^*)}{(2k_Fa)^3}\frac{{\rm cos}(\zeta-\varphi)}
{\zeta^2}\biggr],
\label{e-k}
\end{eqnarray}  
where $N(0)$ and $a$ are the density of states at the Fermi surface 
and the Fe lattice constant, respectively, and $\zeta=2k_F(z+z_o)$.  
The offset $z_o$ arises from the fact that at spacer thickness $z=0$ 
the Fe layers are still the distance between two Fe lattice planes 
apart, which is $z_o \buildrel{>}\over{\sim} a/2$ for the Fe bcc 
structure.  The interlayer coupling $J_1(z)$ calculated from 
Eqs.~{(\ref{e-j}--\ref{e-k})} is shown in Fig.~\ref{f-model}.  Due to 
the exponential damping, only the first AFM minimum of the 
$|2k_F-k_p|$ oscillations is visible and has a strongly asymmetric 
shape, in agreement with experiment (+).  It is important to note that 
any phase shift $\varphi$ of the microscopic RKKY oscillations would 
directly appear as a shift of the position of the minimum by 
$\varphi/|2k_{F}-k_{p}|$.  For the present $a$-Cu$_{65}$Zr$_{35}$ 
samples, where $2k_F$ and $k_p$ differ substantially, one does not 
expect a significant phase shift; however such a phase shift should 
occur when the Nagel--Tauc criterion $2k_F = k_p$ is satisfied 
\cite{haeu92-1,kroh95-1}.  Our measurements show no indication for 
$\varphi \neq 0$.  For $a$-Au$_{60}$Sn$_{40}$ spacers, which were 
studied in Ref.~\cite{fuch97-1}, $2k_F$ and $k_p$ nearly coincide and 
imply a large oscillation period of $\simeq$ 6~nm, rendering the first 
AFM minimum unmeasurably small.  Thus, our model also explains, why no 
AFM coupling was observed in Ref.~\cite{fuch97-1}.

In conclusion, we have measured AFM interlayer coupling across an 
amorphous metallic spacer exhibiting a single pronounced minimum as a 
function of the spacer thickness.  This result is well explained by 
RKKY interaction taking into account structural correlations of the 
amorphous spacer material.  It is proposed that the technique 
presented here provides a direct method to measure a possible phase 
shift of the microscopic RKKY oscillations predicted in 
Ref.~\cite{kroh95-1} for amorphous materials.

Financial support from the Swiss National Science Foundation, the 
Swiss {\em Kommission f\"ur Technologietransfer and Innovation} and 
the DFG SP ''Quasikristalle'' (J.K.) is gratefully acknowledged.


\end{document}